# Income distribution and inequality in India: 2014-19

Anand Sahasranaman[1,2,*] and Nishanth Kumar[3,#]

[1]Division of Science, Division of Social Science, Krea University, Andhra Pradesh 517646, India.

[2]Centre for Complexity Science, Dept of Mathematics, Imperial College London, London SW72AZ, UK.

[3]Dvara Research, Chennai 600113, India.

[*] Corresponding Author. Email: anand.sahasranaman@krea.edu.in

[#] Email: nishanth.k@dvara.com

**Abstract:**

We study the evolution of income in India from 2014-19 and find that while income inequality remains largely consistent over this time, the lower end of the income distribution has experienced significant losses – the bottom ventile shows not only a decline in income share of ~38%, but also negative real average income growth of -4.6% per annum. We further investigate the composition of this part of the distribution using rural and urban splits, and find that even as income shares at the bottom of the urban distribution have increased over time, those at the bottom of the rural distribution have decreased – income share of bottom decile of the rural income distribution declined by ~41%, and real average income growth was at -4.3% per annum from 2014-19. We also empirically confirm that the bottom ventile of the consolidated Indian income distribution is composed primarily of rural incomes, and therefore the decline in real incomes is essentially a rural phenomenon. Studying occupation data of households, we find that the bottom decile of the rural distribution correlates strongly with occupations of small/marginal farmers and agricultural labour, highlighting the increasing economic fragility of such occupations. Using the RGBM model to estimate the nature of reallocation in the Indian income distribution, we find that reallocation has been decreasing from 2015 and even turned negative in 2018, which is in keeping with empirical evidence of real income declines at the bottom of the distribution, and heralds the risk that persistent negative reallocation in the future could result in regressive redistribution of resources from the poor to the rich.

# 1. Introduction

The past 30 years have seen the inexorable rise of income inequality in India (Chancel & Piketty, 2019). This has been caused through a combination of global technological and economic changes as well as structural conditions in the Indian economy since the 1980s (Banerjee & Piketty, 2005; Chancel & Piketty, 2019; Deaton & Dreze, 2002; Kohli, 2012; Dev & Ravi, 2007; Milanovic, 2016).

Immediately after Indian independence in 1947, the state took ownership of the 'commanding heights' of the economy and ensured a progressive taxation structure with very high marginal rates for top incomes, with the explicit goal of curbing elite economic power and driving income convergence - indeed by the early 1970s the top effective marginal tax rate had risen to 97.5% (Banerjee & Piketty, 2005; Chancel & Piketty, 2019; Acharya, 2005). This period, until 1980, saw a sustained decline in income inequality, with the share of the top 10% of income earners reducing from 37% in 1951 to 31% in 1981, and that of the bottom 50% rising from 21% to 24% in the same period (Chancel & Piketty, 2019). This decline in inequality should however be contextualized by the fact that the poverty rate in India remained practically unchanged in this time – from 56% in 1954 to 53% in 1978 (Dutt & Ravallion, 2009). Essentially, low economic growth (3.4% per annum between 1951 and 1980) and high population growth (98% between 1951 and 1981) meant that even with the reallocation of income from the rich to the poor within the distribution, the poverty rate remained persistent at over 50% and the poverty head count doubled (Nagaraj, 1990; Census of India, 2011; Dutt & Ravallion, 2009).

However, since 1980, there has been a progressive dismantling of the socialist architecture of the Indian economy, with enhanced private participation, deregulation of prices, and reduction in tax rates (the top marginal tax rate had declined to 30% in 1998) though still retaining a progressive taxation structure, which has resulted both in increased economic growth and rising income inequality (Banerjee & Piketty, 2005; Kohli, 2012; Chancel & Piketty, 2019; Rodrik & Subramanian, 2004; Basole, 2014). The share of the top 10% of income earners has sharply increased to 56%, and that of the bottom 50% has declined to 15%, as of 2015 (Chancel & Piketty, 2019). However, increased economic growth in the period from 1980 to 2015 (average annual growth rate of 6.05%) has also resulted in a significant decline in the poverty rate to 21% in 2006 (Deaton & Dreze, 2002; Panagariya & More, 2014; Dhongde, 2007; Panagariya & Mukim, 2014; Dutt & Ravallion, 2009; Bhagwati & Panagariya, 2013).

It has been argued that although worsening inequality could be a consequence of economic growth (which yields poverty reduction), that the benefits of this growth are spread across the distribution in India, leaving individuals, on average, better off than before (Bhagwati & Panagariya, 2013). This interpretation is consistent with an income distribution where worsening income shares for those lower in the distribution occurs on account of differential income growths at different points in the distribution – essentially income growth is higher, on average, higher in the distribution, meaning that the lower end of such an income distribution would see lesser than average growth over time and account for a progressively reducing share of total income. In such a distribution, the magnitude of reallocation within the income distribution is progressively declining, but the nature of redistribution is still progressive – from richer to poorer incomes. Recent work has revealed that while this was potentially the mechanism underlying increasing income inequality from 1980 to 2000, it is likely that the nature of redistribution has entered a fundamentally new regime since the early 2000s, where income growth at the bottom of the distribution has not just (on average) been lower, but negative - implying that real incomes in that part of the distribution have actually been declining and the income distribution is essentially diverging (Sahasranaman & Jensen, 2019).

In this work, we study the dynamics of income inequality in India from 2014 to 2019. Using panel data to construct the income distribution, we explore the evolution of income inequality in India as a whole, as well as for the rural and urban separately, to generate an understanding not just of the nature of change in inequality, but also a deeper sense of dynamics at the very bottom of the distribution. Using a stochastic model, we also attempt to quantify the extent and direction of redistribution occurring in the distribution, and reconcile the model's findings with empirical growth incidence curves.

We use the data from the Consumer Pyramids Household Survey (CPHS) published by the Centre for Monitoring Indian Economy. The CPHS is a pan-India panel household survey of roughly 170,000 households collecting monthly data on income, consumption, demographics, assets and borrowing by households. The CPHS dataset creates a geographically representative dataset by sampling one or more Homogeneous Regions (HR) for each state from a set of neighbouring districts that have a similar agro-climatic condition, urbanisation levels, female literacy and family size as per the 2011 Census. The CPHS visits each household in the panel thrice a year (each visit is known as a "wave"), and all household-level is captured at monthly frequency. Using this data, we compute monthly per-capita income by adjusting the total household income reported for each month with the size of the households using a square root equivalence scale (Deaton A., 2003). We use household income as the basis to construct the Indian income distribution by adjusting each income by the appropriate weighting factor (provided by CPHS) to ensure appropriate representation of all household types in the income distribution; and then cumulate these adjusted incomes over each percentile to construct the income distribution. Annual income distributions are obtained by adding the corresponding percentiles in the 12 monthly income distributions.

## 2. Income distribution and inequality

We use a generalization of the Pareto $(80 - 20)$ principle, the $k$-*index*, as a measure of inequality in the income distribution (Banerjee, Chakrabarti, Mitra, & Mutuswami, 2020). Given the Lorenz curve describing the income distribution in a population at a given time, the $k$-index $(k_f)$ reveals that a fraction $(1 - k_f)$ of the population earns a fraction $k_f$ of the total income. From 2014 to 2019, $k_f$ is consistent at $0.65 - 0.66$, indicating that ~35% of the population earned ~65% of income in India through this period. However, when we explore the evolution of incomes of deciles of population, we get a more nuanced picture, revealing that income shares of top 20% declined from 50.7% to 49.1%, while the income share of each decile from the 2$^{nd}$ to the 8$^{th}$ decile showed an increase (Fig. 1a). Given this generally progressive trend, it is the bottom decile which emerges as a concern because it has lost income share in this period – from 1.54% to 1.19%, a decline of ~23% in this five-year period. This decline becomes even more pronounced the deeper we go into the income distribution, with the income share of the bottom 5% (first ventile) declining ~38%, from 0.26% in 2014 to 0.16% in 2019. However, losses in income share over time are not necessarily representative of declines in real income levels; it is possible that certain parts of the income distribution gained share at the expense of others, but that all parts of the distributions experienced absolute increases in real income.

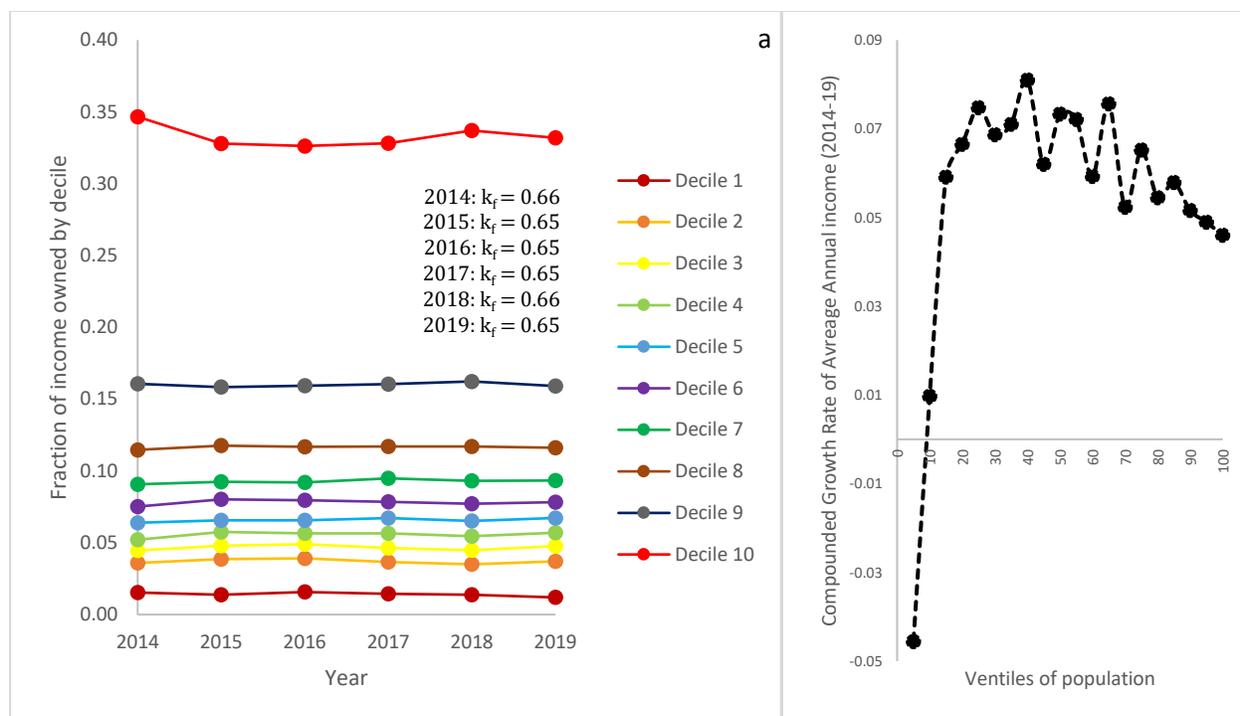

*Figure 1:* **Evolution of the income distribution and Growth Incidence Curve 2014-19.** A: Evolution of income inequality as described the income share of deciles. Income shares of the top income deciles marginally decline and those in the middle increase. The income share of the bottom decile drops by 23%. B: Growth incidence curve (2014-19). The bottom ventile shows negative compounded annual real income growth rates, while the remaining ventiles show positive growth rates. Beyond the bottom 10%, annual growth rates, on average, decline towards the top of the distribution.

In order to explore this, we construct the Growth Incidence Curve (GIC) of average real incomes by ventile for the period 2014-19 and find that the while top 95% of the distribution saw positive compounded annual real income growth from 2014-19 (adjusted for average annual inflation, based on data from the Reserve Bank of India at https://dbie.rbi.org.in/DBIE/dbie.rbi?site=home), the bottom 5% had a decline in real income of -4.56% per annum in this time period (Fig. 1b). Our concerns about declining income shares at the bottom of the income distribution are only exacerbated by this finding that a significant proportion of the lowest incomes experienced negative real income growth. We now attempt to dig deeper into the composition of the bottom of the distribution, by splitting the Indian income distribution into its rural and urban income components.

Before doing so, an important consideration to keep in mind while analysing data from the CPHS survey is that the incomes at the top of the distribution are likely to be underestimated; this is a more general concern with income surveys where the highest incomes are unlikely to participate.

More specifically in this context, we have the Indian income distribution from Chancel and Piketty (2019), who use tax data - a more reliable indicator of top incomes – to find that the income share of the top decile in 2015 was 56%, as against the 32.8% per CPHS data. Therefore, it is possible that the magnitude of income shares of the bottom half is overestimated in this analysis, which potentially makes the situation of the bottom decile of the distribution even more precarious.

## 3. Rural and urban dimensions of income inequality

As with the consolidated income distribution, we find that the $k$-index values for both the rural and urban distributions are between $0.64 - 0.65$ for each year from 2014 to 2019, suggesting that the inequality in the distribution of income remained consistent in both distributions for this period. However, when we assess the change in income shares across parts of the distribution over time, we find that in the urban income distribution, the share of the bottom half increased from 22.8% to 24.4%, while the top half saw a decline in income share from 77.2% to 75.6% (Fig. 2b). In the rural distribution, on the other hand, while we see a decline in income share for the top decile (from 34.2% to 33.6%), the middle part of the distribution - from the 3$^{rd}$ to the 9$^{th}$ decile - experienced increase in income shares, but the bottom two deciles see significant declines, with the income share of the bottom decile declining sharply by ~41%, from 1.27% to 0.76% (Fig. 2a). Given the low levels of income and low income share in this part of the distribution, this represents a drastic reduction. In view of these distinct dynamics in urban and rural distributions, the decline in income share of the bottom decile in the consolidated Indian income distribution appears to be driven primarily by the declines apparent in the rural distribution.

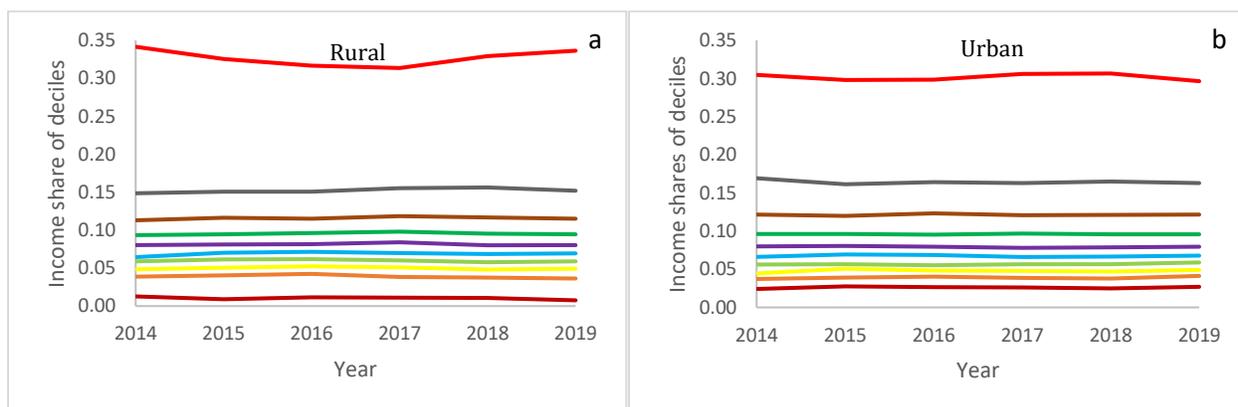

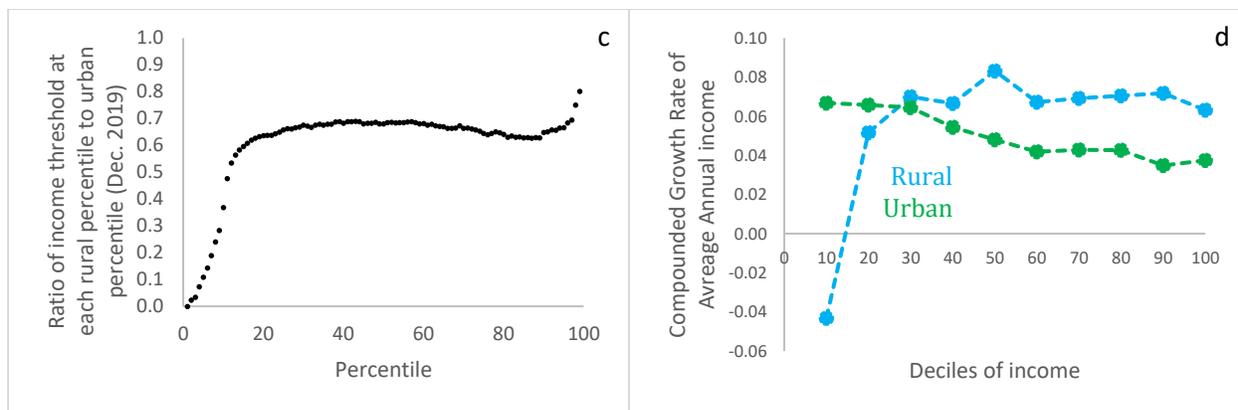

*Figure 2:* **Rural and urban income distributions 2014-19.** A: Income shares of deciles in rural income distribution. While the bottom deciles show loss in income share, the middle deciles show an increase in income share from 2014-19. B: Income share of deciles in urban income distribution. Bottom half of the distribution shows a gain in income share, while top half shows a slight decline. C: Ratio of income threshold per rural percentile to income threshold per corresponding urban percentile in December 2019. Rural incomes are, on average, much lower than urban incomes, and increasingly so as we go towards the bottom of the distribution. D: Growth incidence curves for rural and urban incomes. The bottom decile in the rural distribution shows negative annual income growth rate between 2014-19, but all other deciles in both distributions register positive growth rates.

We find evidence of this when we explore the thresholds of income at each percentile of the distributions – the income threshold of the first percentile of the consolidated Indian income distribution corresponds to the threshold of the first percentile of the rural distribution, and is much lower than first percentile threshold of the urban. Indeed, the first ventile of the consolidated Indian income distribution, which we observe to exhibit negative income growth (Fig. 1b), is composed of the bottom 7 rural income percentiles and only part of the first urban income percentile. This confirms that negative income growth at the bottom of the Indian income distribution is essentially a rural phenomenon. The first decile of the consolidated distribution corresponds to the bottom 14 percentiles of the rural population and only the bottom 3 percentiles of the urban population. Indeed, urban income thresholds are superior to rural thresholds at all points in the distribution, as evinced by the ratio of income thresholds of each percentile in the rural to the urban as in December 2019 (Fig. 2c). This also cautions us against drawing too many inferences from relative performance of equivalent portions in the rural and urban distributions, given these vast discrepancies in income thresholds at corresponding points in the distribution.

We also examine the growth rates of real average income for each decile and find that income growth rate is positive (and progressively declining) for all deciles, pointing to convergence in the urban income distribution, and consistent with an increase in income share for the bottom half of the distribution (Fig. 2d). However, while the top 9 deciles of the rural income distribution exhibit positive real growth, the bottom decile shows negative real growth of -4.32% per annum (Fig. 2d). This result is in agreement with the negative real growth exhibited by the bottom ventile of the consolidated Indian income distribution (Fig. 1b), as the bottom ventile of the Indian distribution is comprised largely of incomes in the bottom 7 percentiles of the rural income distribution (Fig. 2c).

We look into the data to assess the profiles of households at the bottom of the rural income distribution and find that the bottom decile is comprised primarily of small/marginal and organised farmers, as well as agriculture and wage labourers – on average, over 86% of the bottom decile is composed of households with these primary occupations from 2014-19. These occupations, which have experienced declines in real income, therefore comprise the most economically vulnerable workers in the Indian income distribution. Our findings here are in broad agreement with previous findings on rural incomes. Studying rural income inequality between 1993 and 2005, it was found that income from casual labour represented a source of decreasing inequality, meaning that a rise in labour incomes acted as a countervailing force to inequality because it represents the income of those at the bottom of the distribution (Shariff & Azam, 2009). Farm income and salaries, corresponding to higher parts of the rural income distribution, were found to be inequality enhancing sources of income. Analysis of IHDS data also found inequality decreasing effects of income from casual labour and remittances in 2011-12 (Ranganathan, Tripathi, & Rajoriya, 2016).

## 4. Modelling dynamics at the bottom of income distribution

The low and declining income shares of the bottom decile and the bottom ventile, as well as the decline in real income growth in this part of the distribution raise real concerns about the nature of reallocation occurring within the income distribution. Previous econometric modelling work indicates the possibility that the bottom of the income distribution in India has been witnessing negative growth since the early 2000s, and that the overall reallocation within the distribution has turned regressive – there is a perverse transfer of resources from the bottom of the distribution to the top (Sahasranaman & Jensen, 2019).

In order to explore the nature of redistribution occurring in India for the period from 2014-19, we use a simple stochastic model of Geometric Brownian Motion with reallocation (RGBM) to model income dynamics (Berman, Peters, & Adamou, 2017). Empirical work suggests that real world income, expenditure, and wealth distributions are reasonably approximated by lognormal distributions across many national contexts (Chatterjee, Chakrabarti, Ghosh, Chakraborti, & Nandi, 2016; Ghosh, Gangopadhyay, & Basu, 2011; Banerjee, Yakovenko, & Di Matteo, 2006; Drăgulescu & Yakovenko, 2001; Souma, 2001). RGBM models income growth as a multiplicative process described by Geometric Brownian Motion, which yields a widening lognormal distribution over time. However, given the context of real economies, where a number of mechanisms for redistribution are in place (such as taxes, transfers, and public spending), RGBM also incorporates a reallocation parameter ($\tau$) to capture the extend and direction of transfer occurring within the income distribution. Income dynamics in RGBM are described using the following stochastic differential equation:

$$dx_i = x_i(\mu dt + \sigma dW_i) - \tau(x_i - \langle x \rangle_N), \qquad (1)$$

where $dx_i$ is the change in income of $i$ over time $dt$, $\mu$ is the drift and $\sigma$ the volatility of income, $dW_i$ is a Wiener process increment with mean 0 and variance $dt$, $\tau$ is the reallocation parameter, and $\langle x \rangle_N$ is the mean income: $\langle x \rangle_N = \frac{1}{N}\sum_{i=1}^{n} x_i$. The first term of Eq. 1 is the income growth term encompassing growth due to both systemic ($\mu dt$) and idiosyncratic ($\sigma dW_i$) components, and the second is the reallocation term, where the reallocation parameter ($\tau$) is applied to the net difference between individual $i$'s income and the average income of the society. If $\tau > 0$, it is indicative of progressive redistribution, where resources are being reallocated from the top to the bottom of the distribution, which is the reality we would expect in most modern societies; and if $\tau < 0$, over a period of time, the income distribution is divergent and redistribution is occurring from the bottom to the top of the distribution, indicative of a perverse state of economic inequity. $\tau$ is most appropriately understood as a cumulative measure of the overall redistribution occurring in an economy, implicit in the nation's resultant income distribution.

In order to derive $\mu$ and $\sigma$ for the Indian income distribution, we use the previous work of Sahasranaman and Jensen (2019) as the basis for obtaining the values of parameters $\mu$ and $\sigma$ for the Indian income distribution. In that work, using time series data on the Indian income distribution from Chancel and Piketty (2019) and time series of wholesale prices for staple Indian crops and commodities (rich, wheat, and jaggery), it was estimated that $\mu = 0.0231$ and $\sigma = 0.15$ for India's income distribution. Given $\mu$ and $\sigma$, the RGBM algorithm is executed by propagating Eq. 1 over a set of $N$ incomes over $T = 5$ time periods (corresponding to the period from 2014 -19), such that at each time period, the reallocation parameter $\tau(t)$ is obtained by minimizing the distance between the income share of the bottom half of the simulated income distribution ($S_{50\%}^{model}(t)$) and that of the empirically observed Indian income distribution ($S_{50\%}(t)$). In summary, $\tau(t)$ is chosen to minimize $|S_{50\%}^{mod}(t) - S_{50\%}(t)|$ at each time period $t$, resulting in a time series $\tau(t)$ that describes the temporal evolution of both the extent and direction of reallocation apparent in the income distribution. A detailed exposition on the RGBM algorithm is available in Berman, Peters, and Adamou (2017), and for its application to the Indian income distribution in Sahasranaman and Jensen (2019).

Using the RGBM, we model the evolution of the Indian income distribution from 2014 -19, and find reasonable concurrence between modelled results and empirical findings – Fig. 3a describes the modelled incomes for the bottom five deciles (dotted lines) and the corresponding empirical observations (solid lines). Incidentally, the model is able to reasonably simulate the evolution of even the top decile of the distribution, and given that the GBM produces a lognormal distribution, this concurrence supports the concern we had highlighted earlier that the CPHS data is not capturing the power law tail of the Indian income distribution. Given this correspondence between model and observation, we turn our attention to the time series of $\tau(t)$, which describes the time evolution of reallocation within the income distribution (Fig. 3b). $\tau$ is declining from 2015 to 2018, indicating that the extent of reallocation within the distribution is reducing over time and re-distribution is becoming less progressive. Further, for the consolidated (and rural) income distribution, $\tau$ turns negative in 2018 ('17 and '18 for rural) highlighting the risk that continued negative reallocation in the future could result in persistent divergence in the income distribution, yielding a perverse reallocation of resources from the bottom of the distribution to the top. Indeed, the empirical observation of negative real income growth at the bottom of the consolidated and rural income distributions (Figs. 1b, 2d) is already in concurrence with the emergence of negative $\tau$ in these distributions (Fig. 3b). While $\tau$ does emerge positive in 2019 across all distributions, tracking the trend in $\tau$ over a longer period of time will be required to ascertain whether the negative $\tau$ regime is short-lived or not. Longer term perpetuation of negative $\tau$ progressively increases the probability that income growth lower in the distribution turns persistently and deeply negative, calling into question extant policies of economic growth and redistribution in India.

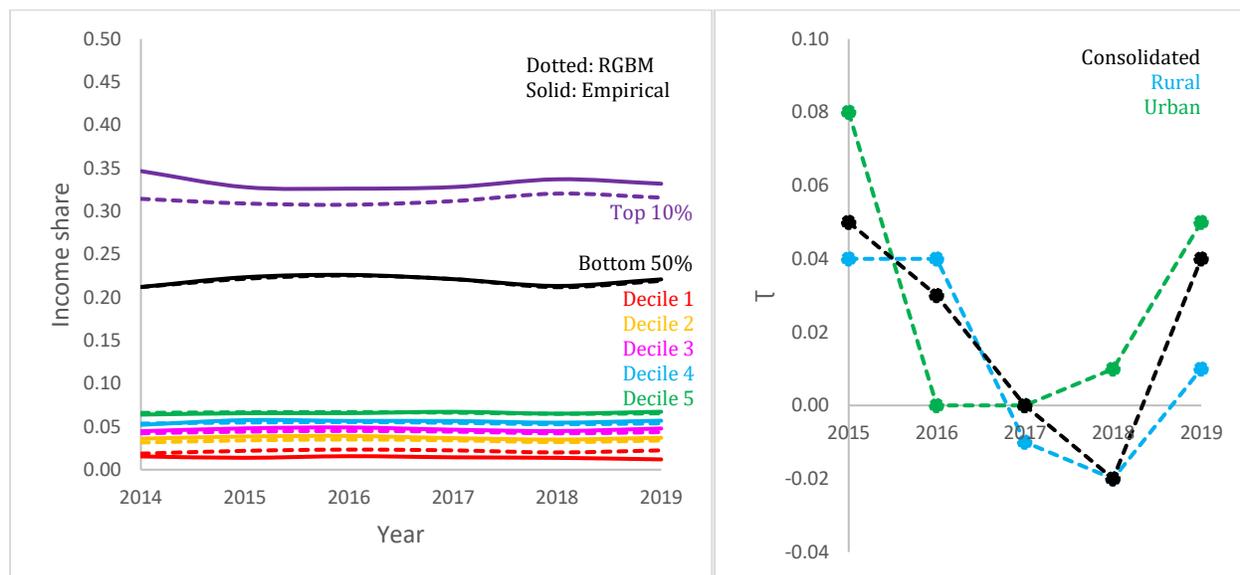

*Figure 3:* **Reallocation under RGBM:** A: Income shares of bottom 5 deciles and top decile (RGBM model v. empirical). Dotted lines represent model outcomes and solid lines empirical data. The income distribution described the RGBM model bears close correspondence with empirical observation. B: Temporal evolution of $\tau(t)$. $\tau$ describes a declining trend across all distributions and even turns negative in for the consolidated and rural income distributions, before recovering to positive territory in 2019.

Previous work has demonstrated that India was likely in a prolonged, decade-long period of negative reallocation beginning in 2002, with a significant proportion of the population at the bottom of the income distribution experiencing negative income growth (Sahasranaman & Jensen, 2019). The growing informalization of the formal workforce in manufacturing and services as well as rising agrarian distress have meant that employment is increasingly characterized by greater insecurity and uncertainty (Vakulabharanam & Motiram, 2011; Vaidyanathan, 2006; Suri, 2006; Mehrotra, 2019). These apparent trends in the emerging context of mechanization and Artificial Intelligence, where in the nature of work is itself expected to change fundamentally over the next decades, mean that the future of employment and income at the bottom of the income distribution is fraught with risk and requires meaningful policy responses.

## 5. Conclusion

We study the income distribution for India from 2014 to 2019 and find that while income inequality remains generally consistent through this period, income shares at the very bottom of the distribution decline substantially. Income shares of the bottom decile and ventile decline by 23% (from 1.54% to 1.19%) and 38% (from 0.26% to 0.16%) respectively. Using Growth Incidence Curves for 2014-19, we find that not only do the income shares of the bottom decline, but that real income growth for the bottom ventile is actually negative – an annual growth rate of -4.6%.

In order to understand the composition of the bottom of the income distribution, we explore rural-urban splits of income distribution, and find that while the bottom half of the urban distribution gains income share from 22.8% to 24.4%, the bottom two deciles of the rural distribution see significant declines, with the income share of the bottom decile declining sharply by 41%, from 1.27% to 0.76%. We also find that the Growth Incidence Curves for the urban and rural distributions reveal that while all deciles in the urban distribution experienced positive real income growth from 2014-19, the bottom decile of the rural distribution experienced negative real income growth for this period, at -4.32% per annum. The bottom ventile of the composite Indian distribution which experienced negative income growth is composed largely of rural incomes. We validate this empirically and find that the bottom ventile of the composite distribution is composed of the first 7 rural percentiles of rural incomes and only part of the first percentile of urban incomes, thus confirming that negative income growth is essentially a rural phenomenon. Using household data for each percentile, we find that the bottom decile of the Indian income distribution is comprised largely of small, marginal, and organised farmers as well as agricultural and wage labour, and it is these workers who are the most economically vulnerable participants in the Indian workforce experiencing declining real incomes.

In order to assess the redistribution occurring within the income distribution, we use the model of Geometric Brownian Motion with reallocation (RGBM) to quantify the nature and extent of reallocation inherent in the income distribution. We find that reallocation rates are declining in all distributions (consolidated, rural, and urban) from 2015 to 2018, and are even negative for the consolidated (in 2018) and rural (2017 and 2018) distributions. This means that the extent of redistribution is decreasing continuously but decline into negative $\tau$ indicates the potential risks of continued negative reallocation – regressive redistribution of resources from the bottom to the top of the distribution. Already, we have evidence of negative real income growth at the bottom of the consolidated and rural distributions from 2014-19. Continually tracking this phenomenon over the next few years will be key to understanding whether the negative reallocation observed is a blip or heralds a longer-term regressive trend in income redistribution.

This fragility of incomes lower in the distribution and decline in real incomes at the bottom is reflective of broader economic trends including informalization of the formal workforce, and agrarian distress. The design of specific policies bearing upon incomes of marginal farmers and wage labourers is therefore an area that requires immediate attention.

**References**

Acharya, S. (2005). Thirty Years of Tax Reform in India. *Economic and Political Weekly*, 40(20): 2061-2070.

Banerjee, A., & Piketty, T. (2005). Top indian incomes, 1922–2000. *WBER*, 19(1): 1-20.

Banerjee, A., Yakovenko, V., & Di Matteo, T. (2006). A study of the personal income distribution in Australia. *Physica A*, 370: 54-59.

Banerjee, S., Chakrabarti, B., Mitra, M., & Mutuswami, S. (2020). On the Kolkata index as a measure of income inequality. *Physica A, 545*, 123178.

Basole, A. (2014). Dynamics of Income Inequality: Insights from World Top Incomes database. *EPW*, 40: 14-17.

Berman, Y., Peters, O., & Adamou, A. (2017). An empirical test of the ergodic hypothesis: Wealth distributions in the United States. *SSRN*.

Bhagwati, J., & Panagariya, A. (2013). Why growth matters: How economic growth in India reduced poverty and the lessons for other developing countries. *Public Affairs*.

Census of India. (2011). Census of India.

Chancel, L., & Piketty, T. (2019). Indian Income Inequality, 1922-2014: From British Raj to Billionaire Raj? *The review of income and wealth*, 65(S1): S33-S62 .

Chatterjee, A., Chakrabarti, A., Ghosh, A., Chakraborti, A., & Nandi, T. (2016). Invariant features of spatial inequality in consumption: The case of India. *Physica A*, 442: 169-181.

Deaton, A. (2003). Health, inequality, and economic development. *Journal of economic literature, 41*(1), 113-158.

Deaton, A., & Dreze, J. (2002). Poverty and inequality in India: A re-examination. *EPW*, 3729-3748.

Dev, S., & Ravi, C. (2007). Poverty and inequality: All-India and states, 1983-2005. *EPW*, 509-521.

Dhongde, S. (2007). Measuring the impact of growth and income distribution on poverty in India. *Journal of Income Distribution*, 16(2): 25-48.

Drăgulescu, A., & Yakovenko, V. (2001). Exponential and power-law probability distributions of wealth and income in the United Kingdom and the United States. *Physica A*, 299: 213–221.

Dutt, G., & Ravallion, M. (2009). Has India's Economic Growth Become More Pro-Poor in the Wake of Economic Reforms? *World Bank Policy Research Working Paper, 5103*.

Ghosh, A., Gangopadhyay, K., & Basu, B. (2011). Consumer expenditure distribution in India, 1983–2007: Evidence of a long Pareto tail. *Physica A*, 390: 83-97.

Kohli, A. (2012). *Poverty Amid Plenty in the New India.* Cambridge University Press.

Lahiri, A. (2020). The great Indian demonitization. *Journal of Economic Perspectives*, 34(1): 55-74.

Mehrotra, S. (2019). *Informal Employment Trends in the Indian Economy: Persistent informality, but growing positive development .* Geneva: ILO.

Milanovic, B. (2016). *Global inequality.* Harvard University Press.

Nagaraj, R. (1990). Growth rate of India's GDP, 1950-51 to 1987-88. *Economic and Political Weekly*, 25(26): 1396-1403.

Panagariya, A., & More, V. (2014). Poverty by social, religious and economic groups in India and its largest states. *Indian Growth and Dev. Rev.*, 7(2): 202–230.

Panagariya, A., & Mukim, M. (2014). A comprehensive analysis of poverty in India. *Asian Dev. Rev.*, 31(1): 1–52.

Ranganathan, T., Tripathi, A., & Rajoriya, B. (2016). Changing sources of income and income inequality among Indian rural households. . *Institute of Economic Growth (IEG), New Delhi*.

Rodrik, D., & Subramanian, A. (2004). The mystery of the Indian growth transition. *NBER Working Paper*, 10376.

Sahasranaman, A. (2020). Long Term Dynamics of Poverty Transitions in India . *SSRN*, 3651204.

Sahasranaman, A., & Jensen, H. (2019). Dynamics of reallocation within India's income distribution. *arXiv*, 1909.04452.

Shariff, A., & Azam, M. (2009). Income Inequality in Rural India: Decomposing the Gini by income sources. *SSRN*, 1433105.

Souma, W. (2001). Universal Structure of the Personal Income Distribution. *Fractals*, 9: 463.

Suri, K. (2006). Political economy of agricultural distress. *EPW*, 1523–1529.

Vaidyanathan, A. (2006). Farmers' suicides and the agrarian crisis. *EPW*, 40: 4009–4013.

Vakulabharanam, V., & Motiram, S. (2011). Political economy of agrarian distress in India since the 1990s. In *Understanding India's New Political Economy* (pp. 117–142). Routledge.

Vyas, M. (2018). Using Fast Frequency Household Survey Data to Estimate the Impact of Demonetisation on Employment. *Review of market integration*, 10(3): 159-183.

Wadhwa, S. (2019). Impact of Demonetization on Household Consumption in India. *Brown University*.